%Paper: hep-lat/9211066
%From: narayana@physics.rutgers.edu (Rajamani Narayanan)
%Date: Mon, 30 Nov 92 18:45:03 EST

\magnification=\magstephalf
\baselineskip=18pt
\line{\hfill RU-92-54}
\bigskip
\centerline{\bf RUNNING COUPLING IN PURE GAUGE THEORIES}
\centerline{\bf USING THE SCHR\"ODINGER FUNCTIONAL}
\medskip
\centerline{ RAJAMANI NARAYANAN }
\smallskip
\centerline{\it Department of Physics and Astronomy, Rutgers University}
\centerline{\it P.O. Box 849, Piscataway, NJ 08855-0849}
\medskip
\centerline{\it ABSTRACT}
\smallskip
 \noindent
{\it
Schr\"odinger functional, the propagation kernel for going from some
field configuration at time $x^0=0$ to some other configuration at
$x^0=T$, is used to define a running coupling, $\bar g^2(L)$, at a
length scale, $L$, in pure gauge theories.
Using a lattice formulation and finite size scaling
techniques, this running coupling is calculated non-perturbatively
in the continuum for
a wide range of $L$, that extends from the perturbative scales to
non-perturbative scales, for the specific case of pure SU(2) gauge
theory.}
\bigskip
If all the parameters of QCD are fixed using low-energy observations,
say, through the hadron spectrum, then the renormalized running coupling,
$\alpha (q)$, is a completely determined function of momentum, $q$.
Different definitions (different renormalization schemes) of
$\alpha (q)$ are related perturbatively at high energies and can be used
as an input for high energy experiments.
In order to compute $\alpha (q)$ from low $q$ to high $q$ in terms
of low energy parameters, it is necessary to use non-perturbative
techniques. One choice for a non-perturbative calculation is the
lattice formulation. To achieve this using a single lattice
would be very difficult because we would need a large lattice
with small lattice spacing to span all momenta from low energies to
high energies. Two recently proposed methods$^1$ to compute the
running coupling using a single lattice face the above mentioned
difficulty.
It is possible to overcome this problem by combining
lattice formulation with finite-size scaling techniques.
This
involves using several small lattices for the simulation rather
than one large lattice.
The
feasibility of this approach was first demonstrated in the case
of O(3) $\sigma$-model in two dimensions$^2$. Recently
the same idea was carried out for the case of pure SU(2) gauge
theory in four dimensions$^{3,4}$.

In pure gauge theories there is only one low energy parameter
which can be taken to be the string tension, $K$. Let $\bar g^2(L)$
be some good definition of the running coupling in a box of size $L$.
$\bar g^2(L)$, so defined, will run with $L=1/q$. Let
$$u=\bar g^2(L)\eqno{(1)}$$
and
$$u'=\bar g^2(sL)\eqno{(2)}$$ where $s$ is a scale factor. Then the function
$$\sigma (s,u)=u',\eqno{(3)}$$
 is the integrated $\beta$-function and can be
computed as a function of $s$ and $u$. We start by choosing an initial
value
$$u_0=\bar g^2(L_0).\eqno{(4)}$$
 Next, we compute
$$u_{i+1}=\sigma(2,u_i);\ \ \ \ \ \ i=0,1,2,\cdots,n-1.\eqno{(5)}$$
This gives us a sequence of values,
$$\bar g^2(L_i)=u_i,\eqno{(6)}$$
 with
$$L_i=2^iL_0.\eqno{(7)}$$
At the terminal point
we compute $L_n\sqrt
{K}$ which fixes all $L_i$ in units of $K$.
In this manner we have the running coupling from the low energy scale
to the high energy scale with the energy scale measured in terms of
$K$, the only low energy parameter.

To compute $\sigma (s,u)$ non-perturbatively we use lattice
regularization and finite-size scaling. On the lattice, we start
by choosing some $u$ and a lattice size $L/a$ where $a$ is the
lattice spacing. The bare coupling $g^2_0$ is tuned so that
$$\bar g^2(L,a,g^2_0)=u.\eqno{(8)}$$
 Then we double the lattice size
and compute
$$\Sigma(2,u,a/L)=\bar g^2(2L,a,g^2_0)\eqno{(9)}$$
 for the same bare
coupling. Then
$$\sigma(2,u)=\lim_{a\rightarrow 0} \Sigma(2,u,a/L)\eqno{(10)}$$ and
is computed by performing the computations for several different $a$
and fixed $L$, i.e; several pairs of lattices $L/a$ and $2L/a$, and
then extrapolating to the limit $a\rightarrow 0$. In SU(2),
$L/a=5,6,7,8,9,10$ were choosen and $\sigma(2,u)$ was computed to a
good accuracy for several values of $u$ $^4$.

To extrapolate $\sigma(2,u)$ from Eq.~(10) using results on lattices
of sizes $L/a=5,6,7,8,9,10$, the $O(a/L)$ correction present in Eqn.
(10) should be small. This will not be true for all choices of the
running coupling.
The running coupling defined through the
Schr\"odinger functional is one choice where the cut off effects can
be small.
The defintion of the running coupling in the continuum is as follows.
Consider the pure gauge theory in a box $L^3T$.
The free energy associated with the Schr\"odinger functional is
$$F[b',b]=-\ln \Biggl\{ {1\over Z} \int D[A]
e^{-S[A]}\Biggr\}\eqno{(11)}$$
In the above equation the functional integral runs over all fields
with
$$A_k(x)=\cases{b^\prime _k(\vec x)&if $x^0=T$\cr b_k(\vec x)&if
$x^0=0$\cr}\eqno{(12)}$$
$$A_\mu(x+\hat k L)=A_\mu(x);\ \ \ \ \ \ k=1,2,3\eqno{(13)}$$
and $S[A]$ is the usual Yang-Mills action. $Z$ in Eq.~(11) is
defined so that $F(0,0)=0$.
The Schr\"odinger
functional describes the propagation from some fixed field
configuration at $x^0=0$ to another fixed field configuration at
$x^0=T$.
The Schr\"odinger functional is a well defined quantity and can be
properly renormalized$^5$. For the case of pure gauge theories no
counter-terms need to be added$^3$. The boundary conditions $b$ and
$b'$ induce a background field $B$ that minimizes $S[A]$. We can
equivalently think of the free energy in Eq.~(11) as $F(B)$.
Let us assume that $B$ depends on one dimensionless parameter
$\theta$,
i.e.; we are varying the boundary conditions with one parameter
$\theta$. Our definition of the running coupling is
$$\bar g^2={F_0^\prime [B]\over F^\prime [B]}\eqno{(14)}$$
where the prime denote differentiation with respect to $\theta$.
If we now choose our boundary conditions, and consequently the
background field to scale with $L$ (we also choose $T=L$), then
the running coupling defined through Eq.~(14) will only depend on $L$
and it will run with $L$. Different choice of background field are
different renormalization schemes. They will all be perturbatively
related at high energies and they can also be related to some
conventional definitions like $g_{\overline{MS}}^2$.

The running coupling defined via the Schr\"odinger functional above can be
carried over to the lattice in the usual manner. The $O(a/L)$ effects
alluded to in the above paragraph will be different for different
choices of background field. We find that an Abelian choice
correspoding to a constant color electric field is one choice where
the $O(a/L)$ effects are small$^3$. The simulations were performed for
this choice of background field and a detailed analysis of the
simulation is presented in Ref. 4.
\medskip
\noindent Acknowledgements:
This research was supported in part by the DOE under grant \#
DE-FG05-90ER40559.
\bigskip
\noindent References:

\item{1.} C. Michael, {\it Liverpool Preprint}
{\bf LTH 279} 1992; A.E. El-Khadra, G. Hockney, A.S. Kronfeld
and P.B. Mackenzie, {\it Fermilab Preprint}
{\bf 91/354-T} 1992.

\item{2.} M. L\"uscher, P. Weisz and U. Wolff,
{\it Nucl. Phys.}
{\bf B359} (1991) 221.

\item{3.} M. L\"uscher, R. Narayanan, P. Weisz and U. Wolff,
 {\it Nucl. Phys.}
{\bf  B384} (1992) 168.

\item{4.} M. L\"uscher, R. Sommer, U. Wolff and P. Weisz,
 {\it DESY Preprint}
{\bf DESY 92-096} July 1992.

\item{5.} K. Symanzik, {\it Nucl. Phys. }
{\bf B190 [FS3]} (1981) 1.

\end